# Tunable kinoform x-ray beam splitter

M. Lebugle,* G. Seniutinas, F. Marschall, V. A. Guzenko, D. Grolimund, and C. David

*Paul Scherrer Institut, CH 5232 Villigen-PSI, Switzerland*
*Corresponding author: maxime.lebugle@psi.ch*

**We demonstrate an x-ray beam splitter with high performances for multi-kilo-electron-volt photons. The device is based on diffraction on kinoform structures, which overcome the limitations of binary diffraction gratings. This beam splitter achieves a dynamical splitting ratio in the range 0-99.1% by tilting the optics and is tunable, here shown in a photon energy range of 7.2-19 keV. High diffraction efficiency of 62.6% together with an extinction ratio of 0.6% is demonstrated at 12.4 keV, with angular separation for the split beam of 0.5 mrad. This device can find applications in beam monitoring at synchrotrons, at x-ray free electron lasers for online diagnostics and beamline multiplexing and, possibly, as key elements for delay lines or ultrashort x-ray pulses manipulation. © 2017 Optical Society of America**

Splitting a beam is a fundamental operation in optics. At visible light wavelengths, the design of beam splitters was achieved already in the early days of optics by exploiting basic mechanisms of linear light-matter interaction. In bulk optics, a plethora of devices is commonly found to realize this operation. A few examples are cube beam splitters based on total internal reflection, devices based on polarization splitting such as the Wollaston or the Glan-Taylor prism, or those for which coatings (metallic or dielectric) are used for partial transmission and reflection at single or multiple interfaces. Beam splitters are key components in a broad range of experiments, for instance, in interferometry, in laser building as output couplers, or in pump-probe experiments, to name a few. In the x-ray realm, however, splitting a beam is not straightforward, particularly in view of obtaining a high splitting efficiency, a high extinction ratio, a dynamical control over the splitting ratio, and an energy tunable device. Various implementations were proposed, for example using single crystals (e.g., silicon or diamond) placed for Bragg reflection [1-3] possibly using an additional beam stop [4], Laue diffraction [5] or a combination of both [6]. Diffractive binary phase gratings can also serve as beam splitters [7], for example, for x-ray monitoring in large-scale facilities [8,9]. At multi-kilo-electron-volt energies, efficient manipulation of light by diffractive optics is particularly difficult due to the long required material length to obtain a significant phase shift (e.g., of $\pi$ radians in phase-shifting zones of a binary phase grating) because of a low refractive index contrast. In this regime, the diffraction angle is in good approximation given by the ratio of the wavelength over the pitch, and diffraction angles that are as large as possible are usually sought. Therefore, high-aspect ratio nanostructures are required, continuously challenging the limits of nanotechnology [10-12].

In this Letter, we present and implement the novel concept of a kinoform x-ray beam splitter, which has a dynamical splitting ratio and can be tuned in photon energy, here demonstrated in the range of 7.2-19 keV. At 12.4 keV, we demonstrated a splitting ratio up to 99.1%, with a splitting efficiency of 62.6%, or -2.0 dB, and an extinction ratio of 0.6%, or -22.2 dB. Other values of the splitting ratio, down to zero, are achieved by using intermediate tilt angles of the beam splitter. The fundamental idea underlying this Letter is to use the 0th and the 1st diffraction orders (DOs) as output ports of the beam splitter, occurring while an incoming x-ray beam impinges on a kinoform grating.

First, to overcome the limitations of binary phase gratings that have a maximum efficiency of 40.5%, our design is based on the kinoform profile. The latter was introduced in context of x-ray lenses and consists of introducing a progressive phase shifting up to $2\pi$ radians in every zone of a zone plate [13-15]. We designed our beam splitter using a kinoform profile; however, here the pitch is constant over the entire aperture, realizing a kinoform diffraction grating. As an asymmetry in the transmission function is introduced, our beam splitter can also be seen as a blazed grating, enhancing the fraction of intensity placed in a given DO. For optimizing the efficiency of the $m$th DO, the optimal phase shift to imprint onto the beam in the thickest part of each grating line is $\Delta\varphi_m = 2\pi m$, where $m$ is an integer. For the 1st DO ($m = 1$), the realization for multi-kilo-electron-volt photons is, however, already a challenge, since the diffractive structure height required to provide a phase shift of $2\pi$ radians is considerable with common materials for x-ray optics such as gold, nickel, or silicon. For increasing the effective structure height, and as the beam splitter is intrinsically one-dimensional, one can tilt the optics [16] (see Fig. 1). When x-rays propagate through the beam splitter in transmission with an angle of incidence $\alpha$, an increase of path length that is proportional to $1/\tan(\alpha)$ is obtained. Simultaneously, the tilted geometry makes it possible to use only binary structures for realizing a continuous phase profile and, thus, avoids resorting to complex fabrication processes such as multi-level stacking [12,17]. To this end, we pattern stripes of triangle nanopillars, which approximate the optimal kinoform phase profile with constant pitch, as depicted in Fig. 1. The effective phase profile equals the sum of the phase profile of each triangular nanopillar experienced by the x-rays, which depends on the real part of the refractive index of the material and the angle of incidence, or the tilt angle. As the beam propagates through this tilted array of triangular nanopillars, the sawtooth phase shift is realized within the accuracy of the fabrication process. Importantly, in contrast to planar refractive lenses [18,19], the aperture is not limited by the structure height and, with current lithographic fabrication techniques, the time for structure patterning does not constrain the area of the lens. Diffraction occurs in the xz-plane defined by the incoming beam and the tilt axis (light gray plane in Fig. 1). The tilt angle is equal to zero when the beam splitter surface is parallel to the beam.

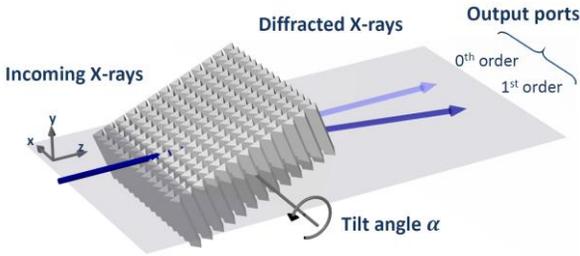

**Fig. 1.** Schematic view of the kinoform x-ray beam splitter.

Secondly, by using the tilt angle as a degree of freedom, one can dynamically set the magnitude of the phase profile to vary the relative intensity distribution between the DOs. This makes it possible to adjust the splitting ratio between the two output ports of the beam splitter, which is similar to adjusting the blazed angle of blazed gratings for given DO and energy. The phase shift can be set from small values to $2\pi m$ radians in the thickest part of each grating line, the upper value being limited by the optics aperture. The set of tilt angles $\alpha_m$ for maximum efficiency of the $m$th DO is

$$\alpha_m = \operatorname{asin}\left(\frac{h\delta g}{m\lambda}\right), \quad (1)$$

with $h$ being the height of the structures, $\delta$ being the real part of the refractive index, $g$ being a factor accounting for the gap separating each lenslet element due to fabrication, and $\lambda$ being the x-ray wavelength. The performances can be assessed by measuring the diffraction efficiency (DE) of the $m$th DO, $d_m(\alpha)$, a function of the tilt angle $\alpha$:

$$d_m(\alpha) = \frac{I_m(\alpha)}{I_{inc}}, \quad (2)$$

where $I_m(\alpha)$ is the intensity of the $m$th DO at the tilt angle $\alpha$, and $I_{inc}$ is the incident intensity. Thus, in the following, the DE includes absorption. As the two output ports of the beam splitter are the 0th and the 1st DO, an ideal device should achieve $d_0(0) = 1$ and $d_1(0) = 0$, a trivial case achieved when the component is placed out of the beam; then $d_0(\alpha_1) = 0$ and $d_1(\alpha_1) = 1$. This corresponds to a lossless material being able to route all the incoming flux towards the 1st DO, with a perfect extinction ratio. All intermediate values of $\alpha < \alpha_1$ allow adjusting the splitting ratio between the output ports. The splitting ratio is

$$s(\alpha) = \frac{d_1(\alpha)}{d_0(\alpha) + d_1(\alpha)}. \quad (3)$$

Finally, using a low-Z element for fabricating an x-ray beam splitter is crucial to keep absorption as low as possible. We target a range of x-ray energies centered on 12.4 keV (wavelength of 1 Å). Therefore, silicon (Si, Z=14) is an interesting candidate, since its K-edge has a characteristic energy of 1.84 keV and has a rather small imaginary part of the refractive index of $3.16\times10^{-8}$ at 12.4 keV. While an x-ray beam with such energy travels through the required length for a $2\pi$ radians phase shift of ~31.3 µm (optimized DE of the 1st DO), a moderate absorption of ~11.7% occurs. Importantly, our beam splitter is also energy tunable, achieved by using a different set of tilt angles $\alpha_m$ [see eq. (1)] for reaching the optimal phase shift of a given DO. The use of Si prevents the use of the device at photon energies close the K-edge while, at high energies, the required extremely shallow angles become limiting. A representative range of energies for high performances is 7.2-19 keV, as further demonstrated.

We fabricated kinoform beam splitters with pitches of 200, 300, and 400 nm (Fig. 2). To achieve Si patterning in triangular nanopillars with sufficient height, we used a metal-assisted chemical etching (MAC-etch) process for etching 10-µm thick Si membranes [orientation (100), boron-doped with conductivity of 1-20 Ω·cm]. This technique transfers a pattern with high fidelity in Si and produces high aspect ratio nanostructures [11,20,21], see Fig. 2(g). E-beam lithography at 100 keV (Vistec EBPG 5000plus, Raith GmbH) was performed on a bilayer of MMA/PMMA resists to expose a triangular stripe negative layout. After development in a solution of isopropanol and water (7:3 in volume), thermal evaporation was used to deposit a 30 nm thick gold film at a rate of 0.25 nm/s. A lift-off procedure was performed in acetone to obtain a negative mask pattern corresponding to the kinoform grating profile. The MAC-etch technique was realized at room temperature with hydrogen peroxide ($H_2O_2$) as an oxidizer and hydrofluoric acid (HF) as an etching agent in a water-based solution. As a result, Si could be etched with the gold pattern acting as a negative mask. The molar concentrations were [HF]=4.93 M and [$H_2O_2$]=0.55 M, resulting in the molar ratio $\rho = [HF]/([HF] + [H_2O_2])$ of 0.9. The reduction reaction of $H_2O_2$ is the rate-limiting step of the redox reaction, which limits the variation in the etch rate as the diffusion lengths for different feature sizes of the catalyst are minimized, as interestingly put forward recently by Chang et al. [11]. The etching rate was about 0.3 µm/min. The samples shown here have a depth of about 5.7 µm, but varies slightly among structures with different pitch due to the influence of the mask on the MAC-etch [21,11]. The resulting aspect ratio is 28.5:1. The gold mask was removed in an aqueous solution of potassium iodide, followed by critical-point drying (Leica EM CPD300 Auto). The support membrane was thinned by about 2 μm from the back side using deep reactive ion etching (Oxford Plasmalab100).

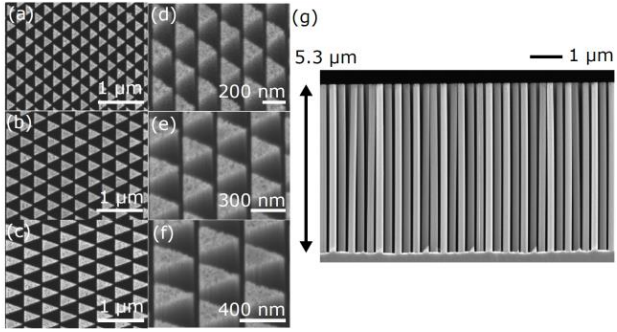

**Fig. 2**. SEM micrographs of kinoform beam splitters on Si membranes with pitches of (a), (d) 200, (b), (e) 300, and (c), (f) 400 nm. (a)-(c) Top view and (d)-(f) tilted view with an angle of 19.5°. (g) SEM micrograph of a cross section of structures with a 250 nm pitch etched into bulk Si, further cleaved for inspection.

X-rays experiments were performed at the microXAS beamline, Swiss Light Source (Paul Scherrer Institute, Switzerland). A fixed-exit double-crystal Si(111) monochromator defined the x-ray energy with a bandwidth of about 2×10$^{-4}$. The beam splitters were aligned using a hexapod (SmarAct) with piezo nanopositioning stages for all six degrees of freedom of translation and rotation. We first characterized at 12.4 keV a kinoform beam splitter with a pitch of 200 nm and diffracting along the vertical dimension. A beam with an aperture of 80 μm x 80 μm was defined by a pair of slits. The detector was at 0.42 m downstream, where a spatial separation of 210 μm between the DOs is found. By vertically scanning a 20-μm slit and measuring the transmitted flux using a photodiode, we obtained the DE $d_m(\alpha)$ of the -8th to the 8th DOs at tilt angles between 4° and 40° (Fig. 3 and inset).

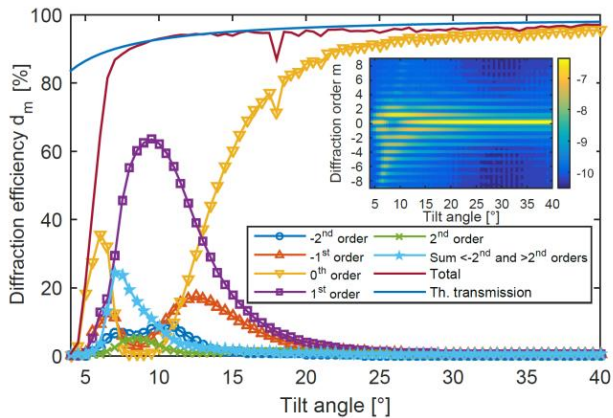

**Fig. 3.** DEs of the -8th to the 8th DOs at 12.4 keV of the beam splitter with a 200 nm pitch versus the tilt angle. The inset shows the raw signal in logarithmic scale obtained by scanning a 20 μm slit across the DOs (vertical axis) versus the tilt angle (horizontal axis). The color bar is the exponent in base 10 of the photodiode current (in A).

As the tilt angle becomes shallower, the 0th DO is attenuated, while the DE of the 1st DO increases, and reaches a maximum of 62.6% at a tilt angle of $\alpha_1^{exp} = 9°$, theoretically expected at 7.4° using Eq. (1). This difference probably arises from the fact that, in Eq. (1), a lossless material is assumed, shifting the expected angles towards shallower values. At the tilt angle optimum for the 1st DO, our tilt geometry permits a six-fold increase of the effective aspect ratio of the nanostructures. Other maxima of higher DOs, such as the -1st, -2nd, and 2nd also occur and may be attributed to either the imperfect shape of the kinoform elements or secondary maxima of the DE as a function of the material length passed through, i.e., not occurring at the optimum phase shift of $\Delta\varphi_m = 2\pi m$. At an angle of 7°, a significant fraction of 24.6 % is found in the sum of DEs of the ±3rd, ±4th, ±5th, ±6th, ±7th, and ±8th DOs (see the light blue line with star markers). This reveals the higher sensitivity of high DOs to small imperfections of the kinoform shape. The sum of all measured DEs (dark red line) agrees well with the calculated overall transmission (dark blue line). A deviation is observed at 18° that may come from the (400) Si Bragg reflection (expected at 21.6°), which would indicate a miscut of the crystal of about 3.6°. In view of our application, the DEs of the beam splitter with a 200 nm pitch are $d_0(\alpha_1^{exp}) = 0.6\%$ and $d_1(\alpha_1^{exp}) = 62.6\%$, for the 0th and the 1st DOs, respectively. This leads to a maximum splitting ratio of $s(\alpha_1^{exp}) = 99.1\%$. At an angle of 12.9°, the device acts as a 50:50 beam splitter, with DEs of the 0th and the 1st DOs nearly identical and equal to 35.1%. Table 1 and Fig. 4 summarize similar measurements of the DEs with larger pitches of 300 and 400 nm, also expressed in decibels.

**Table 1. Optimum of DEs of the 0th and 1st DOs at 12.4 keV of Beam Splitters with Pitches of 200 nm, 300 nm, and 400 nm**

| Pitch | 200 nm | 300 nm | 400 nm |
|---|---|---|---|
| $d_0(\alpha_1^{exp})$ | 0.6% | 0.6% | 0.8% |
| | -22.2 dB | -22.2 dB | -21.0 dB |
| $d_1(\alpha_1^{exp})$ | 62.6% | 71.4% | 74.8% |
| | -2.0 dB | -1.5 dB | -1.3 dB |
| Splitting angle | 0.5 mrad | 0.33 mrad | 0.25 mrad |

The maximum DE increases with pitch, being as high as 74.8% (-1.3 dB) for a pitch of 400 nm, as the shape of larger nanostructures is better controlled [see Figs. 2(c) and 2(f)]. The beam splitter extinction ratio remains strong, less than 0.8% (-21.0 dB), and 50:50 splitting is possible for all devices. Using the beam splitter with a pitch of 300 nm, images of both output ports on a scintillator screen were obtained along with the splitting ratio $s(\alpha)$ and $1 - s(\alpha)$; see Figs. 4(d) and 4(e). We also realized efficiency measurements from 7.2 keV to 19 keV (Table 2). Similar performances were obtained while keeping an aperture of 80 μm x 80 μm. A slight increase in DE of the 1st DO with energy is observed, as the silicon absorption coefficient decreases accordingly.

**Table 2. Optimum of DEs of the 0th and 1st DOs at Several Energies of a Beam Splitter with a Pitch of 200 nm**

| Energy | 7.2 keV | 16 keV | 19 keV |
|---|---|---|---|
| $d_0(\alpha_1^{exp})$ | 1.0% | 1.2 % | 1.9% |
| | -20.2 dB | -19.2 dB | -17.2 dB |
| $d_1(\alpha_1^{exp})$ | 51.3% | 62.1% | 64.1% |
| | -2.9 dB | -2.1 dB | -1.9 dB |

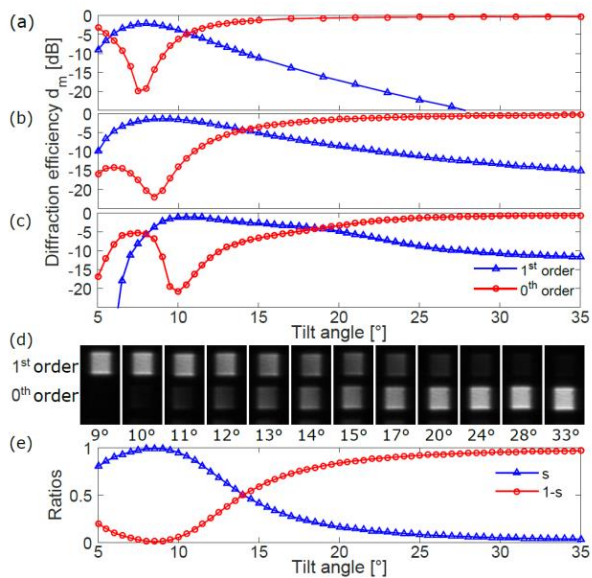

**Fig. 4.** DEs of the 0th and 1st DOs at 12.4 keV of beam splitters with pitches of (a) 200, (b) 300, and (c) 400 nm, versus the tilt angle. (d) DOs imaged on a scintillator screen using a 300 nm pitch beam splitter, with (e) the corresponding splitting ratios (see text).

Our beam splitter could be used for x-ray beam diagnostics requiring variable flux, either at synchrotrons or x-ray free-electron lasers (X-FELs). At X-FELs, multiplexing experiments could be enabled by routing part of the photon flux to a secondary setup. To increase the angular separation between beams, Bragg reflection onto a thin Si crystal can be used. As the splitting angle (see Table 1) is larger than the corresponding rocking curve width, one could tune only one split beam for Bragg reflection to obtain a larger split angle. One application could be a split-and-delay line for ultrashort x-ray pulses, where two pulses are generated and recombined with varied delay and relative intensities, to perform pump-probe or double-pump experiments in a non-collinear or collinear geometry. Eventually, the nanostructures of the kinoform beam splitter can form a Fresnel bi-prism, i.e., two thin prisms joined at their base. In such a device, the left portion of the wavefront is deflected right, and vice-versa for the right portion, creating a zone of interference as two virtual sources exist. This can be used for source size or coherence length measurements [22].

Summarizing, we devised a tunable beam splitter for multi-kilo-electron-volt x-ray wavelengths with high efficiency and extinction ratio, and whose splitting ratio can be dynamically adjusted. We foresee the use of such beam splitters for beam diagnostics in large-scale facilities, and possibly as key element in ultrafast x-ray optics.

**Funding.** Horizon 2020 Framework Programme (H2020) (654360 NFFA-Europe).

**Acknowledgment.** The x-ray experiments were performed at the microXAS beamline of the Swiss Light Source, Paul Scherrer Institute, Villigen, Switzerland. The authors thank Florian Dworkowski for fruitful discussions about kinoform optics. They are also grateful to Lucia Romano and Dario Marty for their help while developing the MAC-Etch process, and to Dario Ferreira Sanchez for his dedicated support during the measurements at microXAS.